\documentclass[final]{svjour2}
\usepackage{graphicx}
\usepackage{rotating}
\usepackage{amssymb}
\usepackage{mathptmx}
\usepackage[numbers]{natbib}
\makeatletter
\journalname{Journal of Low Temperature Physics}

\bibpunct{}{}{,}{s}{}{,}

\begin{document}

\newcommand{\hdblarrow}{H\makebox[0.9ex][l]{$\downdownarrows$}-}
\title{Critical velocities in two-component superfluid Bose
gases}

\author{L.Yu. Kravchenko \and D.V. Fil}

\institute{Institute for Single Crystals, National Academy of
Sciences of Ukraine, Lenin av.60, Kharkov 61001, Ukraine\\
Tel.:+38(057)3410492\\ Fax:+38(057)3409343\\
\email{fil@isc.kharkov.ua}}

\date{XX.XX.2007}

\maketitle


\begin{abstract}

On the ground of the Landau criterion we study the behavior of
critical velocities in a superfluid two-component Bose gas. It is
found that under motion of the components with different
velocities the velocity of each component should not be lower than
a minimum phase velocity of elementary excitations ($s_-$). The
Landau criterion yields a relation between the critical velocities
of the components ($\mathbf{v}_{c1}$, $\mathbf{v}_{c2}$). The
velocity of one or even both components may exceed $s_-$. The
maximum value of the critical velocity of a given component can be
reached when the other component does not move. The approach is
generalized for a two-component condensate confined in a
cylindrical harmonic potential.

\end{abstract}

\section{Introduction}

At present considerable attention is given to the study of
two-component Bose-Einstein condensates. The progress in cooling
and trapping of atomic rarefied gases allows to obtain such
condensates experimentally \cite{Rb2, Rb1,Na, K-Rb,Cs-Li}. From
the fundamental point of view, they are considered as objects in
which some cosmological and astrophysical processes can be
modelled \cite{fi,al,ba}. In two component systems the superfluid
components may flow with different velocities. In such a situation
an unusual (nondissipative) kind of the drag effect takes place
\cite{ab}. It makes possible to create a controlled phase
difference between two Bose-condensates placed in a two-well
potential, and to observe effects \cite{fil}, similar to ones that
occur in superconductive systems with Josephson contacts in
magnetic fields. In this article we study a related problem,
namely, the critical velocities in a two component superfluid
system with components flowing with different velocities.

\section{The Landau criterium for the two-component system}

According to the Landau criterion, the critical velocity in a
single-component condensate is determined by the expression
 \begin{equation}\label{1}
    v_{\mathrm{c}} = \min \left( \frac{E_0(k)}{\hbar k} \right),
\end{equation}
where  $E_0(k)$ is the excitation spectrum in an immovable
condensate and $k$ is the wave number. The velocity is given in
the frame of reference connected with walls or obstacles. Eq.
(\ref{1}) can be applied to a two-component condensate only in the
case when both components move with the same velocity.

The Landau criterium can be reformulated as the requirement of
positivity of energies of elementary excitations in the frame of
reference, connected with  walls or obstacles.  These energies
depend on the velocities of the components ${\bf v}_1$ and ${\bf
v}_2$ (in the same frame of reference), and the Landau criterium
should yield some joint condition on ${\bf v}_1$ and ${\bf v}_2$.

To find the energy of elementary excitations we  use the
Gross-Pitaevskii equation for the two-component system
\begin{equation}\label{5}
i \hbar \frac{\partial\psi_j}{\partial t} = - \frac{\hbar^2}{2m_j}
\nabla^2 \psi_j + \gamma_j |\psi_j|^2 \psi_j + \gamma_{12}
|\psi_{(3-j)}|^2 \psi_j,\ (j=1,2),
\end{equation}
where $\psi_j$ are the wave functions of the components, $m_j$ are
the masses of the particles, $\gamma_i$, $\gamma_{12}$ are the
interaction constants ($\gamma_{j} = 4 \pi \hbar^2 a_{jj} / m_j$,
$\gamma_{12} = 2 \pi \hbar^2 (m_1 + m_2) a_{12} / (m_1 m_2)$ where
$a_{ik}$ are the scattering lengths).

The wave function of the component can be represented as the sum
of the stationary part and the fluctuating part $\psi_j =
\psi_{0j} + \delta\psi_j$, where $\delta \psi_j \ll \psi_{0j}$.
The stationary part of the condensate wave function can be
presented in the form:
\begin{equation}\label{6}
  \begin{array}{r}\displaystyle
   \psi_{0j} ( \mathbf{r},t ) =  \sqrt{n_j} \, e^{i \varphi_j (\mathbf{r})} e^{-
   \frac{i \mu_j t}{\hbar}},
 \end{array}
\end{equation}
where $\mu_{j} = \displaystyle \frac{m_j \mathbf{v}_j^2}{2} +
\gamma_{j} \, n_j + \gamma_{12} \, n_{3-j}$ are the chemical
potentials of the components. The gradients of the phases
$\varphi_j$ are connected with the superfluid velocities by the
relation
 $\mathbf{v}_j = \displaystyle \frac{\hbar}{m_j} \nabla \varphi_j$.

The fluctuating part can be written as
\begin{equation}\label{14}
  \begin{array}{r}\displaystyle
\delta \psi_j (\mathbf{r},t) = e^{- \frac{i \mu_j t}{\hbar}} e^{i
\varphi_j (\mathbf{r})} \left[u_j  e^{i({\bf k\cdot r}-\omega t)}
+ v_j^{\ast} e^{-i({\bf k\cdot r}- \omega t)} \right].
 \end{array}
\end{equation}
The substitution of (\ref{14}) into the linearized version of
(\ref{5}) leads to a system of equations for $u-v$ coefficients,
whose determinant gives the dispersion equation for the spectrum
of elementary excitations
\begin{equation}\label{16}
[E_1^2 - (E - \hbar \mathbf{v}_1\cdot \mathbf{k})^2] [E_2^2 - (E -
\hbar \mathbf{v}_2\cdot \mathbf{k})^2] - 4 \varepsilon_1
\varepsilon_2 \gamma_{12}^2 n_1 n_2 = 0,
\end{equation}
where $E_j=\sqrt{ \varepsilon_j(\varepsilon_j+2\gamma_{jj} n_j)}$
is the Bogolyubov spectrum for the $j$-component (in the absence
of interaction between the components), and $\varepsilon_j=\hbar^2
k^2/2 m_j$.

For ${\bf v}_1 = {\bf v}_2 = {\bf v}$ the equation (\ref{16})
yields
\begin{equation}\label{18}
E_{\pm} = \sqrt{\frac{E_1^2 + E_2^2}{2} \pm \sqrt{\frac{(E_1^2 -
E_2^2)^2}{4} + 4 \gamma_{12}^2 \varepsilon_1 \varepsilon_2 n_1
n_2}}+\hbar \mathbf{k}\cdot \mathbf{v}.
\end{equation}
In this study we assume that the condition of stability of a
two-component condensate relative to phase separation is fulfilled
($\gamma_1 \gamma_2 - \gamma_{12}^2 > 0$).

The requirement of positivity of (\ref{18}) at all ${\bf k}$ is
equivalent to the condition (\ref{1}).  This condition yields the
following expression for the critical velocity
\begin{equation}\label{27} v_{\mathrm{c}} =s_-=
\frac{1}{\sqrt{2}}\sqrt{s_1^2+s_2^2-\sqrt{(s_1^2-s_2^2)^2+4s_1^2
s_2^2 \frac{\gamma_{12}^2}{\gamma_1\gamma_2}}},
\end{equation}
where $s_j=\sqrt{\gamma_j n_j/m_j}$ are the bare velocities of the
sound modes for the components (in the absence of interaction
between the components).

In a general case ${\bf v}_1 \neq {\bf v}_2$ the Landau criterium
requires the existence of two positive solutions of Eq. (\ref{16})
at all ${\bf k}$. This requirement is equivalent to the following
two inequalities:
\begin{equation}\label{30}
[E_1^2 - (\hbar \mathbf{v}_1\cdot \mathbf{k})^2] [E_2^2 - (\hbar
\mathbf{v}_2\cdot \mathbf{k})^2] - 4 \varepsilon_1 \varepsilon_2
\gamma_{12}^2 n_1 n_2 > 0,
\end{equation}
\begin{equation}\label{31}
E_1^2 > (\hbar \mathbf{v}_1\cdot \mathbf{k})^2 \quad \textrm{(or }
E_2^2
> (\hbar \mathbf{v}_2\cdot \mathbf{k})^2 \textrm{ ) }.
\end{equation}
The critical values of ${\bf v}_1$ and ${\bf v}_2$ correspond to
the case, when the inequality (\ref{30}) turnes into the equality
at least for one ${\bf k}$. In the case considered the sufficient
condition for fulfilling the inequalities (\ref{30}), (\ref{31})
for all ${\bf k}$ is their fulfillment at $k\to 0$ for all
directions of ${\bf k}$, complanar to ${\bf v}_1$ and ${\bf v}_2$.
Therefore, the inequalities (\ref{30}), (\ref{31}) can be replaced
with the system of inequalities
\begin{equation}\label{20}
\left( s_1^2-v_1^2 \cos^2 \alpha  \right) \left(s_2^2-v_2^2
\cos^2(\theta-\alpha)\right) - \frac{\gamma_{12}^2}{\gamma_1
\gamma_2} s_1^2 s_2^2 > 0, \ s_1^2>v_1^2 \cos^2 \alpha
\end{equation}
(where $\theta$ is an angle between ${\bf v}_1$ and ${\bf v}_2$,
and $\alpha$ is an angle between ${\bf k}$ and ${\bf v}_1$), that
should be fulfilled for all $\alpha$. The results of the analysis
of (\ref{20}) is given in Fig.1
\begin{figure}
\begin{center}
\includegraphics[%
  width=0.75\linewidth,
  keepaspectratio]{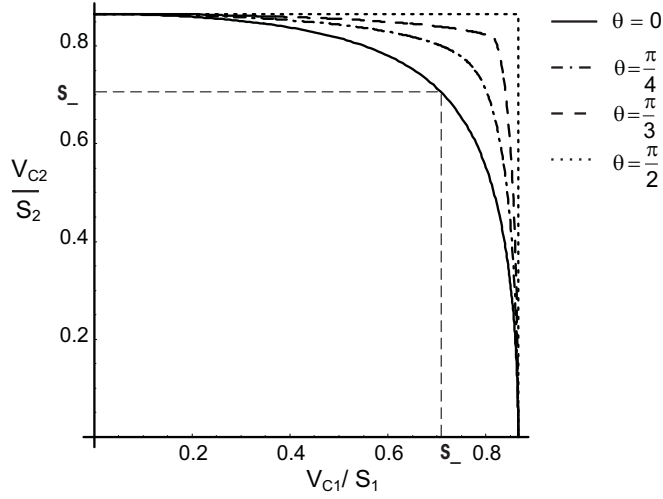}
\end{center}
\caption{Connection between the critical velocities at different
$\theta$ (for the parameters $\gamma_{12}=\sqrt{\gamma_1
\gamma_2}/2$). The shown value $s_-$ corresponds to the case
$s_1=s_2$. } \label{f2}
\end{figure}

If the components move  in the same direction their critical
velocities are related by the equation
\begin{equation}\label{112}
    (s_1^2-
    v_{\mathrm{c1}}^2)(s_2^2-v_{\mathrm{c2}}^2)=\frac{\gamma_{12}^2}{\gamma_1\gamma_2}s_1^2
    s_2^2
\end{equation}
with the additional condition  $v_{c1}<s_1$. According to
(\ref{112}) only for ${\bf v}_1={\bf v}_2$ the critical velocity
coincides with the velocity of the lowest hydrodynamic mode $s_-$.
In a general case one of the velocities may exceed $s_-$ (the
second velocity  should be less than $s_-$). If one of the
components does not move the velocity of the other component may
reach the maximum value
\begin{equation}\label{113}
    v_{\mathrm{c}j,\mathrm{max}}=  s_j \sqrt{1 - \frac{\gamma_{12}^2}{\gamma_1
    \gamma_2}}.
\end{equation}

One can see from Fig.1 that at $\theta\ne 0, \pi$ both components
may move with velocities that exceed $s_-$. Under the motion of
the components in mutually perpendicular directions ($\theta = \pi
/2$) the velocities can simultaneously reach the maximum critical
values (\ref{113}).

Here we do not consider the possibility of excitation of vortices.
Therefore, strictly speaking, our analysis yields only the upper
bound for the critical velocities. Nevertheless, in a number of
situations the estimation for the critical velocities presented in
this paper is justified completely. For example, this occurs when
the superfluid flows past an obstacle with a small (less than the
healing length) linear size \cite{st,max,car}.

\section{Critical velocities in the two-component condensate, confined
in a cylindrical harmonic potential}

In Bose gases confined in optical or magnetic traps surface
excitations have the minimum phase velocity \cite{fedich}. In such
systems the process of generating of vortices is connected with
the excitation of surface modes, and the critical velocity
coincides with the phase velocity of the lowest surface mode
\cite{anglin}.

Let us study the critical velocities for a two-component Bose gas,
confined in a harmonic cylindrical potential $V (r) = m \omega_0^2
(x^2+y^2)/ 2$. We will consider the case when superfluid flows are
directed along $z$. For simplicity we assume $n_1(r)=n_2(r)=n(r)$
($r$ - radial coordinate), $m_1=m_2=m$, $\gamma_1=\gamma_2=\gamma$
and $0<\gamma_{12}<\gamma$. Let us consider the system whose
Thomas- Fermi radius $R_{\mathrm{TF}}=[2(\gamma+\gamma_{12})n_0/m
\omega_0^2]^{1/2}$ is much large than the oscillator length of the
trap. To find the spectrum of elementary excitations we pass from
the Gross -Pitaevskii equation to the linearized system of
hydrodynamic equations for the densities $n_j (\mathbf{r}, t)$ and
the velocities $\mathbf{v}_j (\mathbf{r},t)$ of the components:
\begin{equation}\label{29}
  \begin{array}{l}\displaystyle \vspace*{0.3cm}
\frac{\partial \delta n_j}{\partial t} + \nabla (n \mathbf{\delta
v}_j + \mathbf{v}_{0j} \delta n_j) = 0,
\\  \displaystyle m \frac{\partial \delta \mathbf{v}_j}{\partial t} +
\nabla (\gamma \delta n_j +\gamma_{12} \delta n_{3 - j} + m
\mathbf{v}_{0j}\cdot  \mathbf{\delta v}_j) = 0,
 \end{array}
\end{equation}
where $n=n_0(1-r^2/R_{\mathrm{TF}}^2)$ is the equilibrium  density
of the components ($n_0$ is the density in the center of trap),
$\mathbf{v}_{0j}=(0,0,v_j)$ are their superfluid velocities and
$\delta n_j$ and $\mathbf{\delta v}_j$ are the fluctuations of
these values.

The  modes we are interested in are localized near the surface and
the problem considered can be reduced to  the problem for the
spectrum of excitations in a Bose gas in a linear potential
\cite{sh, pt, fsh01}. With this simplification we obtain the
following dispersion equation
\begin{equation}\label{24}
\left[ (\omega - k v_1)^2 - \frac{2\gamma n_0 k}{m
R_{\mathrm{TF}}}\right] \left[ (\omega - k v_2)^2 - \frac{2\gamma
n_0 k}{m R_{\mathrm{TF}}}\right] -  \frac{4 \gamma_{12}^2 n_0^2
k^2}{m^2 R_{\mathrm{TF}}^2}= 0.
\end{equation}
The region of the applicability of (\ref{24}) is bounded from
above by the condition $k\lesssim k_{\mathrm{m}}$ where
$k_{\mathrm{m}}$ is the wave vector  for which the contributions
of kinetic and potential energies to the excitation energy become
comparable. The value of $k_{\mathrm{m}}$ can be estimated by
equating the kinetic energy of the particles $\hbar^2k^2/2m$ to
the energy of the lowest mode (in the hydrodynamic limit)
$E_{0,-}= \hbar [2 (\gamma - \gamma_{12}) n_0 k / m
R_{\mathrm{TF}}]^{1/2}$ what gives $k_m=2\sqrt[3]{\displaystyle
\frac{m n_0(\gamma-\gamma_{12})}{\hbar^2 R_{\mathrm{TF}}}}$. At
the point $k \approx k_{\mathrm{m}}$ the dependence of the
excitation energy on $k$ has a bend  and the critical velocity can
be estimated by substituting of $k=k_{\mathrm{m}}$ into the
equation (\ref{24}).

As it follows from  (\ref{24}) for $v_1=v_2$ the critical velocity
is the minimum phase velocity of the lowest surface mode
\begin{equation}\label{a1}
    v_{\mathrm{c}}=s_{\mathrm{sf}}=\left(\frac{2\gamma
    n}{m
    R_{\mathrm{TF}}k_{\mathrm{m}}}\right)^{1/2}\sqrt{1-\frac{\gamma_{12}}{\gamma}}.
\end{equation}
If only one component moves, the critical velocity reaches the
value
\begin{equation}\label{a2}
    v_{\mathrm{c,max}}=\left(\frac{2\gamma n}{m R_{\mathrm{TF}} k_{\mathrm{m}}}\right)^{1/2}
    \sqrt{1-\frac{\gamma_{12}^2}{\gamma^2}}=s_{\mathrm{sf}}\sqrt{1+\frac{\gamma_{12}}{\gamma}}.
\end{equation}
Thus, the highest velocity can be reached when only one component
flows.

\section{Conclusions}

In conclusion, let us discuss a number of possibilities to observe
the predicted behavior. One of them is to create  barriers not
penetrable for one of the components. If in the absence of the
barriers the components move with equal velocities the appearance
of such barriers may increase of the critical velocity. The motion
of the components with different velocities (and in different
directions) can also be realized in systems, in which the
components are separated spatially, for example, in a bilayer
geometry (in that case a long-range interaction between the
components is required for observing the effect predicted). The
effect is also may be observed  in multilayer condensates of
electron-hole pairs which can appear in semiconducting
heterostructures with an even ($>2$) number of two-dimensional
electron layers \cite{jl}. In electron-hole condensates the
critical velocities  can be easily measured since they are
proportional to the value of critical currents.


\end{document}